\documentclass[12pt,a4paper]{article}
\usepackage{amsmath}
\usepackage[utf8]{inputenc}
\usepackage[left=2cm,right=2cm,top=2cm,bottom=3cm]{geometry}
\usepackage{ifthen}
\usepackage{xspace}
\usepackage{mciteplus}
\usepackage[hidelinks,breaklinks=true]{hyperref}
\usepackage{cleveref}
\usepackage{upgreek}
\usepackage{caption} 
\usepackage{overpic}
\usepackage{ifthen}
\usepackage{lineno}

\usepackage{graphicx}  
\newboolean{inbibliography}
\setboolean{inbibliography}{false} 
\newboolean{articletitles}
\setboolean{articletitles}{true} 
\usepackage{cite} 
    
\title{\LARGE{\textbf {Anomalous measurements:\\recent results deepen flavour puzzle}}}
\author{Davide Lancierini$^{1}$, Dan Moise$^2$\\
{\normalsize $^1$\textit{Physik-Institut, Universit\"at Z\"urich, Z\"urich, Switzerland}}\\
{\normalsize $^2$\textit{Imperial College London, London, United Kingdom}}
}
\date{21$^{st}$ June 2021}

\newcommand{\btosll}{\ensuremath{b \to s \ell^+ \ell^-}\xspace}
\newcommand{\BKll}{\mbox{\ensuremath{B^\pm \rightarrow K^\pm \ell^+ \ell^-}}\xspace}

\newcommand{\Bsmumu}{\mbox{\ensuremath{B_s^0 \rightarrow \mu^+ \mu^-}}\xspace}
\newcommand{\Bsmumugamma}{\mbox{\ensuremath{B^{0}_{s}\rightarrow \mu^{+}\mu^{-}\gamma}}\xspace}
\newcommand{\Bsmumug}{\mbox{\ensuremath{B_s^0 \rightarrow \mu^+ \mu^- \gamma}}\xspace}
\newcommand{\Bdmumu}{\mbox{\ensuremath{B^0 \rightarrow \mu^+ \mu^-}}\xspace}
\newcommand{\Bsdmumu}{\mbox{\ensuremath{B_{(s)}^0 \rightarrow \mu^+ \mu^-}}\xspace}
\newcommand{\RK}{{\ensuremath{R_K}}\xspace}
\newcommand{\RKst}{{\ensuremath{R_{K^{*}}}}\xspace}
\newcommand{\JPsi}{{\ensuremath{J/\psi}}\xspace}
\newcommand{\PsiS}{{\ensuremath{\psi(2S)}}\xspace}
\newcommand{\rJPsi}{{\ensuremath{r_{\JPsi}}}\xspace}
\newcommand{\RPsiS}{{\ensuremath{R_{\PsiS}}}\xspace}
\newcommand{\BKJPsill}{\mbox{\ensuremath{B^\pm \rightarrow K^\pm J/\psi(\ell^+ \ell^-)}}\xspace}
\newcommand{\BKmumu}{\mbox{\ensuremath{B^\pm \rightarrow K^\pm \mu^+ \mu^-}}\xspace}
\newcommand{\BKee}{\mbox{\ensuremath{B^\pm \rightarrow K^\pm e^+ e^-}}\xspace}
\newcommand{\BKstmumu}{\mbox{\ensuremath{B \rightarrow K^{*} \mu^+ \mu^-}}\xspace}

\newcommand{\qsq}{\ensuremath{q^2}\xspace}
\def\RKvalue {\ensuremath{0.846\,^{+\,0.042}_{-\,0.039}~{\rm(stat.) }\,^{+\,0.013}_{-\,0.012}~{\rm(syst.) }}}
\def\significance {{3.1}\xspace}
\def\BFBsmumuvalue{\ensuremath{3.09\,^{+\,0.46}_{-\,0.43}~{\rm(stat.) }\,^{+\,0.15}_{-\,0.11}~{\rm(syst.) }}}
\def\BFbdlimit{\ensuremath{2.6 \times 10^{-10} \textrm{ at } 95\% \textrm{ C.L.}}}
\def\BFbsgammalimit{\ensuremath{2.0 \times 10^{-10} \textrm{ at } 95\% \textrm{ C.L.}}}
\newcommand{\qsqRange}{\mbox{\ensuremath{\qsq\in(1.1\,{\rm GeV}^2,\;6.0\,{\rm GeV}^2)}}\xspace}

\begin{document}
\maketitle

\vspace{1cm}

\renewcommand{\baselinestretch}{1.25} \normalsize

\begin{abstract}
The crown jewel of particle physics, the Standard Model (SM), has withstood numerous experimental trials. However, there are still some observations it cannot explain. 
Examples such as dark matter and the matter-antimatter imbalance in the Universe come to mind.
The SM may be extended, by including additional particles and interactions, so as to explain such phenomena. These new particles and interactions are collectively referred to as ``new physics'' (NP), and the results covered by this article provide a promising lead for their discovery.
\end{abstract}

\vspace{3cm}

\noindent Invited contribution to the newsletter of the CERN EP Department, June 2021\\
(reformatted with minimal changes for submission to arXiv) \\
\href{https://ep-news.web.cern.ch/content/anomalous-measurements-recent-lhcb-results-deepen-flavour-puzzle}{\texttt{https://ep-news.web.cern.ch/node/3265}}

\newpage

\section{Flavour anomalies}

In the past decade, a pattern has been emerging in the study of $b$-quark decays with leptons in the final state. They are collectively referred to as ``flavour anomalies'', and they typically feature tensions at the level of $2$--$3$ standard deviations between experimental results and SM predictions. As such, individual anomalies are not sufficiently significant to claim discovery of NP. However, the anomalies are often treated collectively in an Effective Field Theory (EFT) framework, whereby short distance contributions are separated from their long distance counterparts. This is similar to the $4$-point approximation used to describe beta decay, where the process is observed at long enough distances to regard the $W$ boson propagator as point-like.  
The EFT approach leads to the formulation of an effective Lagrangian, $\mathcal L_{\rm eff}=\sum_i C_iO_i$, where large energy-scale effects are encoded in the so-called Wilson coefficients $C_i$, and small energy-scale contributions are accounted for by the Wilson operators $O_i$. Given the absence of direct discoveries at the LHC, NP is assumed to be characterised by large (above TeV) energy scales. The anomalous flavour observables could then be impacted indirectly, since NP particles may manifest themselves virtually. From an EFT prespective, this would lead to values of the $C_i$ coefficients that differ from the SM values.

\begin{figure}[!h]
\begin{center}
\begin{overpic}[width=.37\textwidth]{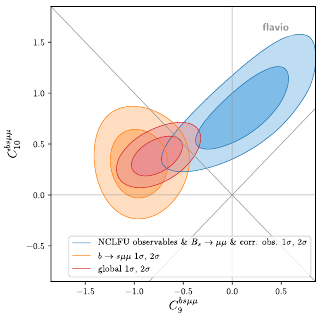}
\put(70.5,37.7){$\star$ {\scriptsize SM}}
\end{overpic}
\begin{overpic}[width=.56\textwidth]{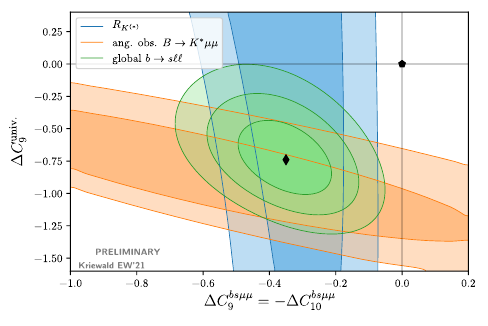}
\end{overpic}
\end{center}
\vspace{-.75cm}
\captionof{figure}{Examples of EFT fits to anomalous flavour observables~\cite{Altmannshofer:2021qrr,Alguero:2021anc,Hati:2020cyn}, under different NP scenarios.
Blue confidence regions show values preferred by flavour anomalies that are theoretically cleaner compared to other anomalies, whose preferred values are depicted in orange.
Confidence regions preferred by all considered flavour anomalies are shown in red and green.}
\label{fig:globalFits}
\end{figure}

Two examples of EFT analyses of flavour anomalies are shown in~\Cref{fig:globalFits}~\cite{Altmannshofer:2021qrr,Alguero:2021anc,Hati:2020cyn}. In the plot on the left-hand side, the $1\,\sigma$ and $2\,\sigma$ confidence levels for the best-fit point for the considered anomalies are shown in red. On the right-hand side, the best-fit $1\,\sigma$, $2\,\sigma$, and $3\,\sigma$ confidence levels are shown in green. 
The two analyses make different assumptions on the presence of NP in the coefficients $C_i$ (as shown by the axes), however they both find that the flavour anomalies prefer $C_i$ values that are significantly different from the ones predicted by the SM (the origin in each plot).

\section{Clean observables}

\begin{figure}[!h]
   \begin{center}
      \includegraphics[width=.48\textwidth]{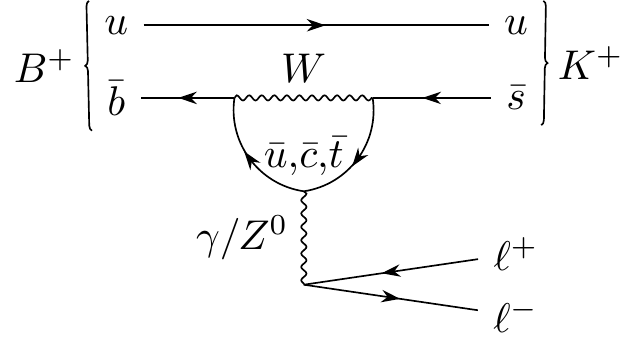}
      \includegraphics[width=.48\textwidth]{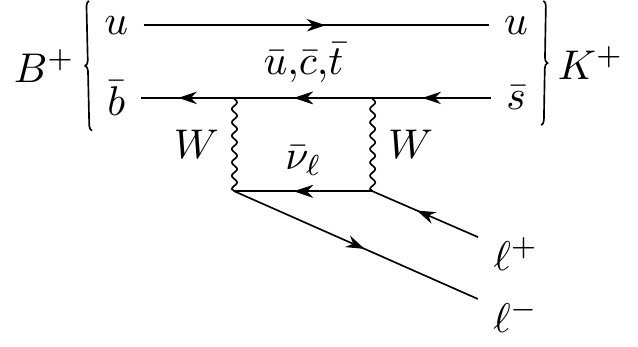}
      \includegraphics[width=.48\textwidth]{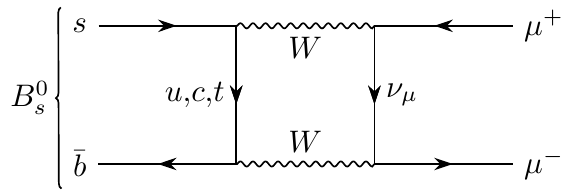}
   \end{center}
\vspace{-.75cm}
   \captionof{figure}{Examples of lowest-order Feynman diagrams allowed in the SM for \BKll (top) and \Bsmumu (bottom). Since FCNCs are forbidden at tree-level, each of these diagrams contains a loop.}\label{fig:FDbtosllSM}
\end{figure}

There are several types of anomalous flavour observables, such as differential branching fractions and angular coefficients. They are predicted in the SM with various degrees of precision, and so particularly important are the flavour anomalies that are theoretically clean. Recently, at the Electroweak session of the Moriond 2021 conference, the LHCb collaboration has presented the most precise individual measurements to date of two such theoretically-clean anomalous observables. The first one is the ratio \mbox{$\RK=\mathcal B(B^\pm\to K^\pm\mu^+\mu^-)/\mathcal B(B^\pm\to K^\pm e^+e^-)$}~\cite{LHCb-PAPER-2021-004}. The second one is the branching fraction of \Bsmumu, including a search for \Bdmumu and \Bsmumug~\cite{LHCb-PAPER-2021-007,LHCb-PAPER-2021-008}. The processes involved are examples of Flavour Changing Neutral Currents (FCNCs), which are forbidden at tree-level in the SM. As a result, the leading-order Feynman diagrams contain loops, as exemplified by~\Cref{fig:FDbtosllSM}. The loops introduce additional vertices that make these processes rare.
The branching fraction of \Bsmumu is predicted by the SM with sub-percent precision, thanks to the purely-leptonic final state: \mbox{$\mathcal B(\Bsmumu)=(3.66\pm0.14)\times10^{-9}$}~\cite{Beneke:2019slt}. The two \BKll processes upon which \RK is built have branching fractions of $\mathcal O(10^{-6})$~\cite{PDG2020}. They differ only by the leptons in the final state, which means \RK can be predicted accurately by virtue of Lepton Flavour Universality (LFU). This is an accidental symmetry of the SM, whereby the different lepton flavours couple in the same way to vector bosons. As a result, \RK is predicted to be approximately $1$, with small deviations induced by phase space differences and QED corrections~\cite{GinoRKPrediction1Percent,Isidori:2020acz}.

\begin{figure}[!h]
\begin{center}
\begin{overpic}[width=.535\textwidth]{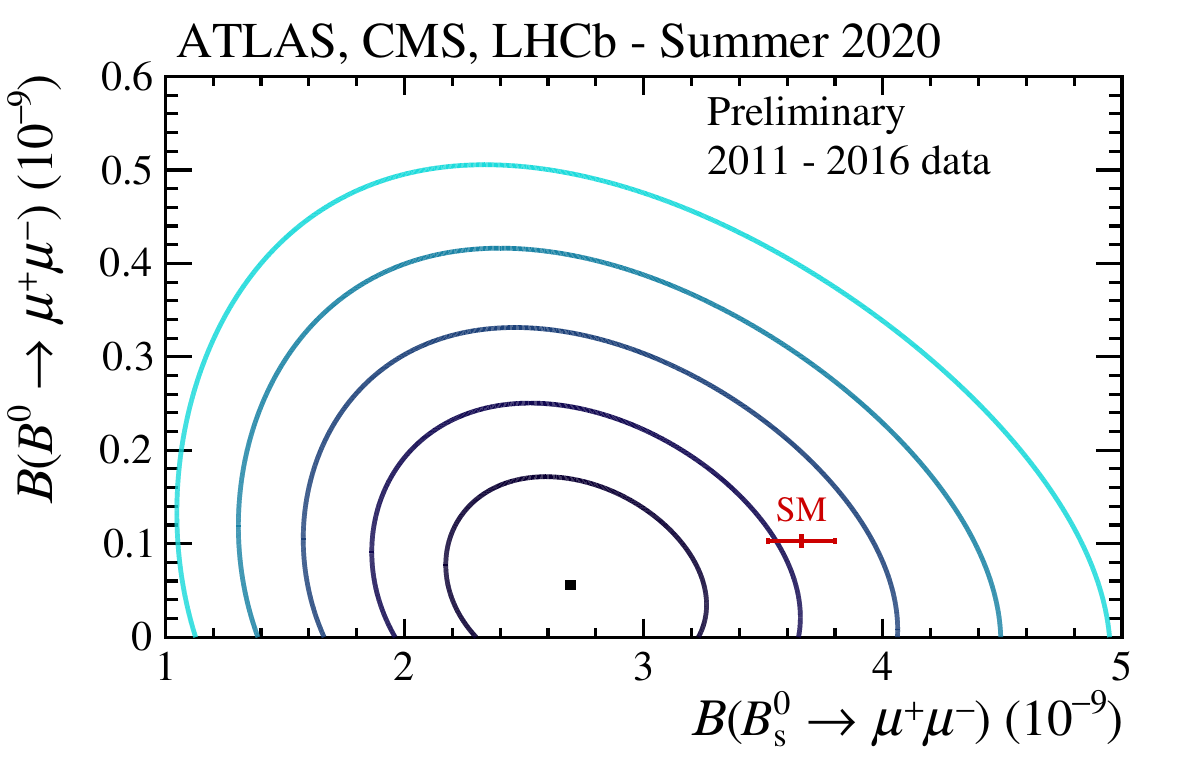}\end{overpic}
\begin{overpic}[width=.455\textwidth]{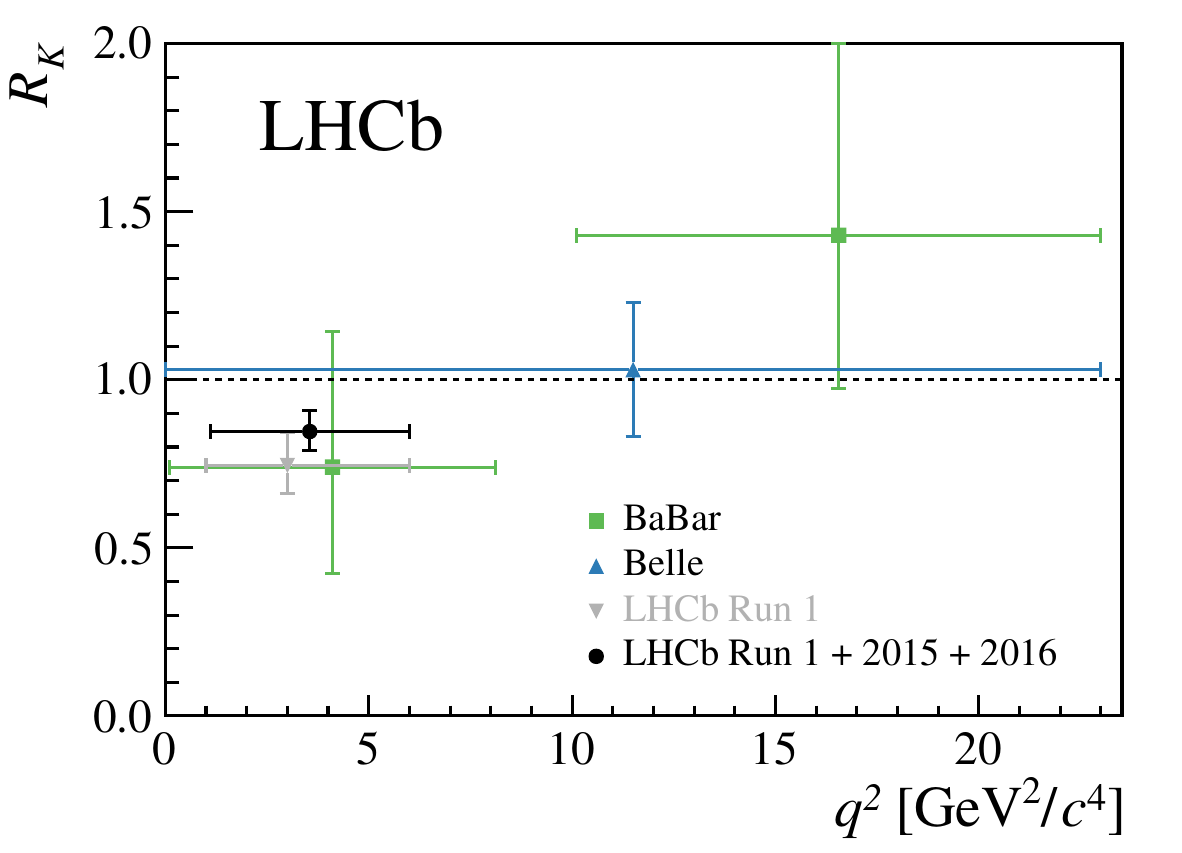}\end{overpic}
\end{center}
\vspace{-.75cm}
\captionof{figure}{Experimental status of \Bsdmumu (left) and \RK (right) before the Moriond 2021 conference. Shown on the left are $1$--$5\,\sigma$ confidence regions (shades of blue) for the combination~\cite{LHCb-CONF-2020-002} of \Bsmumu and \Bdmumu measurements from ATLAS~\cite{Aaboud:2018mst}, CMS~\cite{Sirunyan:2019xdu}, and LHCb~\cite{LHCB-PAPER-2017-001}, alongside the SM prediction (red). Shown on the right are results on \RK from LHCb~\cite{LHCb-PAPER-2019-009,LHCb-PAPER-2014-024} (black and grey), BaBar~\cite{RK_babar} (green), and Belle~\cite{RK_belle} (blue); the latter has since been updated~\cite{Abdesselam:2019lab}. The SM prediction is shown as a dashed line at $\RK=1$.
}
\label{fig:resultsOld}
\end{figure}

Despite the unambiguous SM predictions, experimental results on \Bsdmumu and \RK were exhibiting tensions above $2\,\sigma$ before the Moriond 2021 conference, as shown in~\Cref{fig:resultsOld}. The LHCb measurements shown here used $5\,{\rm fb}^{-1}$ of proton-proton collision data, corresponding to approximately half of the currently available dataset. Given the bigger picture in the context of the flavour anomalies, updates to \Bsdmumu and \RK based on the full LHCb data set are crucial to better understanding whether there is NP in rare $b$-decays.

\section{Measurement of \boldmath{\RK}}

Highly anticipated updates to the \RK and \Bsdmumu measurements, now including  the full dataset of $9\,{\rm fb}^{-1}$ of proton-proton collisions collected by LHCb, were presented at the Moriond 2021 conference. At the core of the \RK measurement is the comparison of selection efficiencies between electrons and muons. 
The masses of the two leptons differ by two orders of mangitude, which leads to them interacting differently with the detector.
Muons traverse the LHCb detector almost undisturbed, before finally being stopped by the muon stations. However, electrons are subject to significant loss of energy to bremsstrahlung radiation, leading to worse momentum and mass resolution compared to muons. In order to suppress detection effects that are systematically different between the two lepton flavours, \RK is measured as a double ratio:
\begin{align}\label{eq:expRK}
\begin{split}
    \RK & =\left.\frac{N(K^{\pm}\mu\mu)}{N(K^{\pm}ee)}\frac{\varepsilon(K^{\pm}ee)}{\varepsilon(K^{\pm}\mu\mu)} \middle/\underbrace{\frac{N(K^{\pm}J/\psi(\mu\mu))}{N(K^{\pm}J/\psi(ee))}\frac{\varepsilon(K^{\pm}J/\psi(ee))}{\varepsilon(K^{\pm}J/\psi(\mu\mu))}}_{r_{J/\psi}}\right.\,, 
\end{split}
\end{align}
where $N(X)$ and $\varepsilon(X)$ represent respectively the yields and efficiencies of selecting the decay of a $B^{+}$ meson into X. 
The $J/\psi$ modes upon which the single ratio \rJPsi is built are used to calibrate efficiencies and conduct cross-checks. These modes are characterised by high statistics, and are known to respect LFU~\cite{PDG2020}. In addition, the \BKJPsill (control) channels are kinematically similar to the \BKll (rare) channels, and so the data are selected with identical requirements for both. The only exceptions are the cut on the dilepton invariant mass squared, \qsq, and the reconstructed $B$ mass. The control data are selected using a \qsq window around the \JPsi resonance, whilst the \BKll candidates are required to have \qsqRange. This \qsq window is chosen to prevent contamination from resonances such as $\phi(1020)$ (at low \qsq) and \JPsi (at high \qsq).

Several cross-checks are performed to ensure the selection efficiencies are well understood. Among these checks are the ratios \rJPsi and \RPsiS. Like \RK, the latter is a double ratio with respect to \rJPsi, the difference being that the numerator comes from the \PsiS resonance. The ratios are found to be $\rJPsi=0.981 \pm 0.020\,({\rm stat+syst})$, and $\RPsiS=0.997 \pm 0.011\,({\rm stat+syst})$. Both are compatible with the LFU expectation of $1$.
This confirms that the efficiencies are valid across \qsq, and that systematic effects successfully cancel out in the double ratio. 

\begin{figure}[!h]
   \begin{center}
      \includegraphics[width=.48\textwidth]{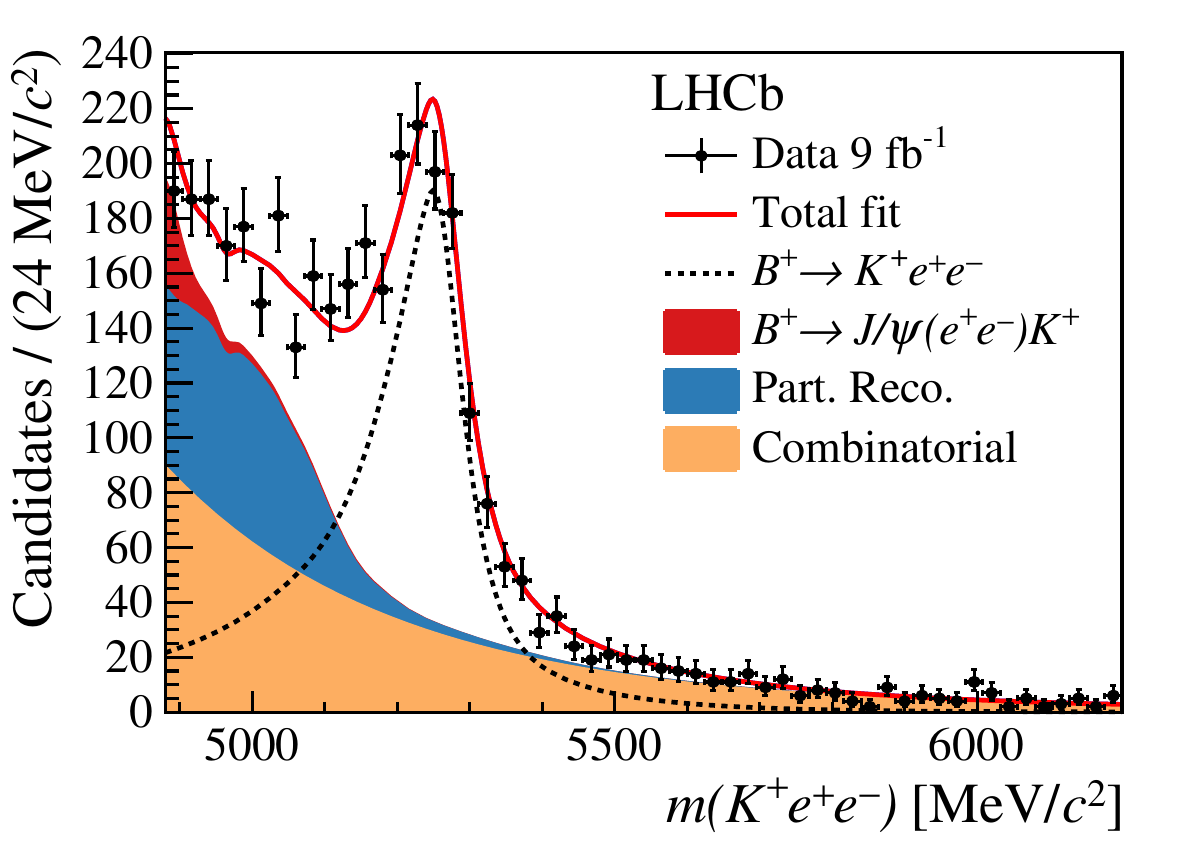}
      \includegraphics[width=.48\textwidth]{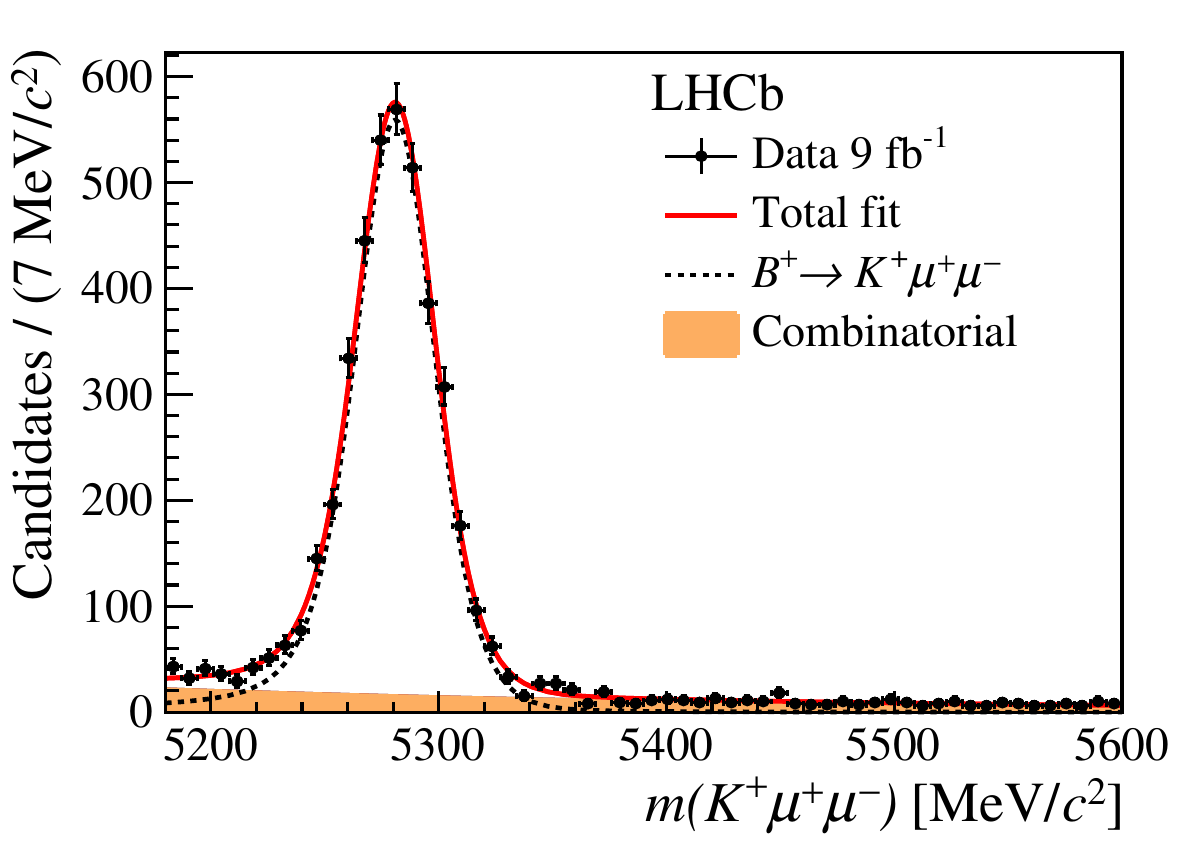}
   \end{center}
\vspace{-.75cm}
   \captionof{figure}{Fit projections for \BKee (left) and \BKmumu (right) candidates. The resolutions and background components differ as a result of the different behaviour of electrons and muons as they traverse the LHCb detector.}
   \label{fig:raremodefits}
\end{figure}

The value of \RK is extracted from an unbinned maximum likelihood fit to the selected \BKee and \BKmumu data. The different mass resolutions lead to different background components in the fit model, as shown in~\Cref{fig:raremodefits}. The result for \RK~\cite{LHCb-PAPER-2021-004} is:
\begin{align}
    \RK = \RKvalue\,.
\end{align}
As expected, the uncertainty on the result is dominated by the statistical component, rather than the systematic one. The dominant systematic effect is the choice of the fit model, which contributes by around $1\%$. By comparison, effects induced by the calculation of efficiencies are reduced to the permille level by the double ratio.

\section{\texorpdfstring{Branching fractions of \boldmath\Bsdmumu$\!(\gamma)$}{Branching fractions of B(s) --> mumu(g)}}

\begin{figure}[!h]
   \begin{center}
      \includegraphics[width=.48\textwidth]{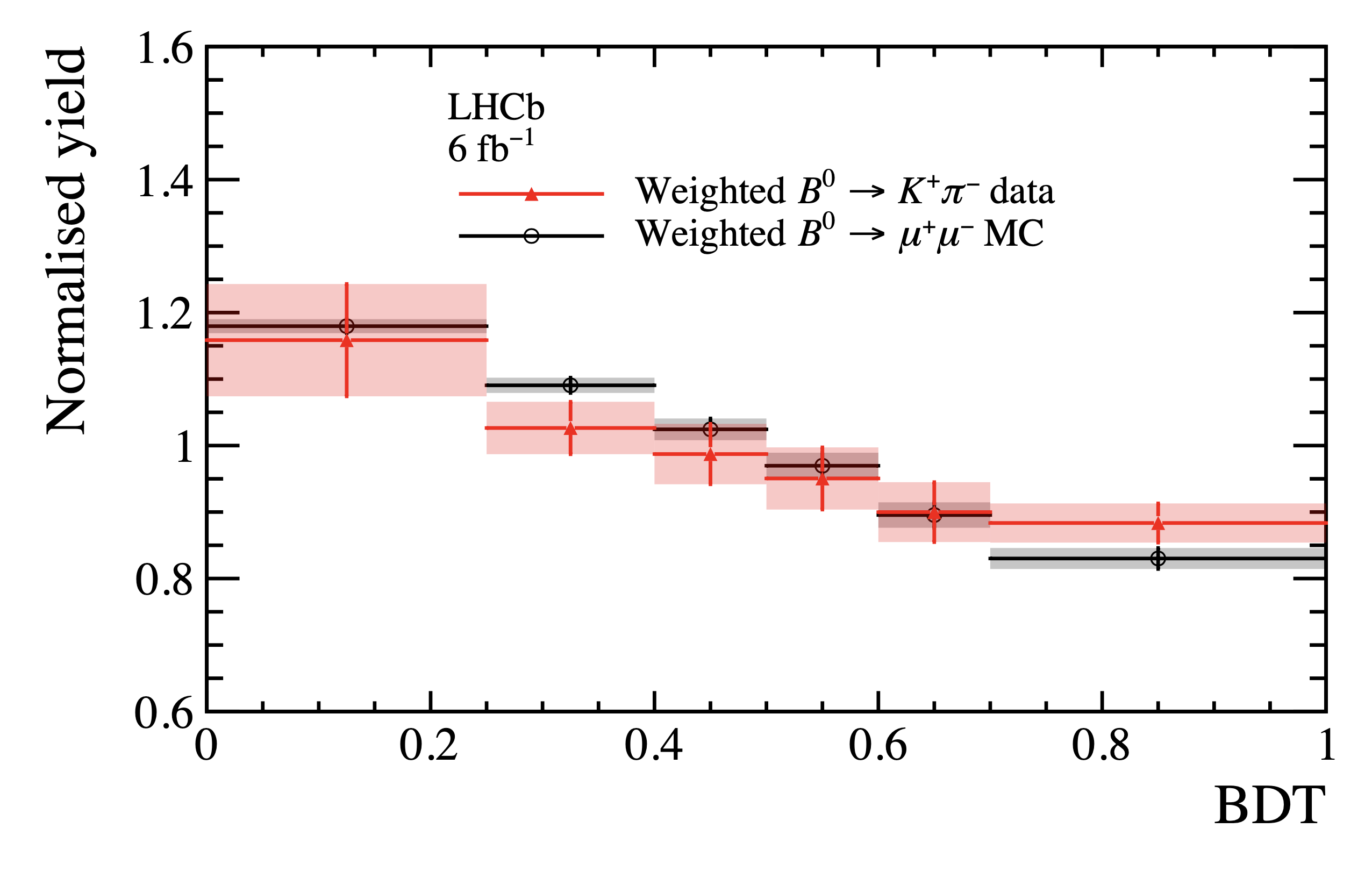}
      \includegraphics[width=.45\textwidth]{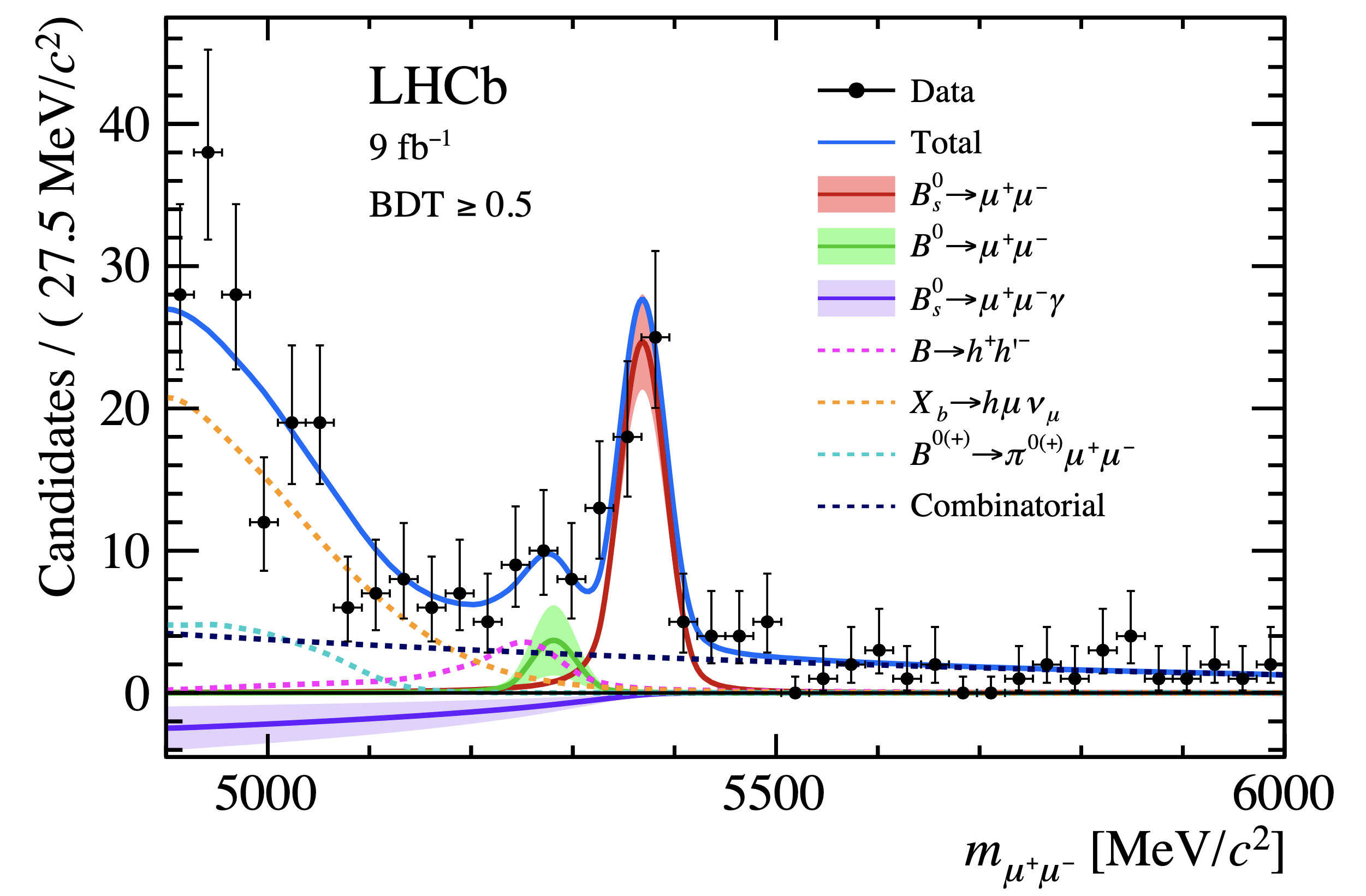}
   \end{center}
\vspace{-.75cm}
   \captionof{figure}{Left: expected distribution of the relative signal yield, as a function of the BDT score.
     Right: invariant-mass distribution of \Bsdmumu candidates with BDT score above $0.5$. The total fit model is shown in blue, alongside the individual components that represent signal and background.}
   \label{fig:bsmumufits}
\end{figure}

Thanks to the excellent mass resolution of muons, the signature of \Bsdmumu decays is very clean. It consists of a pair of oppositely charged muons that have an invariant mass around mass of the $B_{(s)}$, and that form a vertex that's significantly displaced from the interaction point. The branching fraction of the signal decays are extracted from an unbinned maximum likelihood fit to the dimuon invariant mass and are measured relatively to two normalisation channels $B^{\pm}\rightarrow \JPsi K^{\pm} $ and $B^{0} \rightarrow K^{+} \pi^{-}$ which both share similarities with the signal. The former is characterised by similar particle identification and trigger performance, and the latter has similar kinematics. The branching fraction is written as:
\begin{align}
\mathcal{B}(\Bsdmumu(\gamma)) =\mathcal{B}_{{\rm norm}} \frac{f_{\rm sig}}{f_{\rm norm}} \frac{N_{\rm sig}}{N_{\rm norm}} \frac{\varepsilon_{\rm sig}}{\varepsilon_{\rm norm}}\,,
\end{align}
where $N_{\rm sig}$ and $N_{\rm norm}$ are the yields of the signal and normalisation channels, respectively. The corresponding efficiencies and fragmentation fractions are denoted by $\varepsilon_{\rm norm(sig)}$ and $f_{\rm norm(sig)}$ respectively. LHCb provided a measurement for the latter in~\cite{Aaij:2021nyr}. In order to maximise sensitivity to the signal, a boosted decision tree (BDT) is used to separate signal and background.
The fit for the \Bsdmumu yields is performed simultaneously on subsets of the data separated by BDT output. The left-hand side plot of \Cref{fig:bsmumufits} shows the expected relative yield of \Bsdmumu, in bins of BDT score, as estimated using two different methods. The red distribution uses a BDT calibration based on the control channel $B^{0} \rightarrow K^{+} \pi^{-}$, as done in the previous LHCb analysis~\cite{LHCB-PAPER-2017-001}. The black histogram uses calibrated \Bsdmumu simulation. The two methods yield compatible results, and the new BDT calibration method significantly improves the precision on the expected relative yield. This is important because it directly improves the total uncertainty of the branching fraction measurement.

The branching fraction of \Bsmumu is obtained from a fit to the invariant mass of the dimuon system, which is shown on the right-hand side of~\Cref{fig:bsmumufits}. In addition to the \Bsmumu signal, the fit model contains contributions from \Bdmumu and \Bsmumug. These two components are compatible with the background-only hypothesis, and so upper limits are set on their corresponding branching fractions. In summary, the results are:
\begin{align}\label{eq:expBsmumu}
\begin{split}
    \mathcal{B}(\Bsmumu) &= \BFBsmumuvalue\,,\\
    \mathcal{B}(\Bdmumu) &< \BFbdlimit\,,\,\text{and}\\
    \mathcal{B}(\Bsmumugamma)_{m_{\mu\mu}>4.9\,\textrm{GeV}} &< \BFbsgammalimit\,.
\end{split}
\end{align}
Like \RK, the measurement of $\mathcal{B}(\Bsmumu)$ is dominated by statistics. The systematic uncertainty is largely due to the uncertainty on $f_{s}/f_{d}$.
The measured values are compatible with the previous results, as well as with the SM predictions, at the $1\,\sigma$ level.

\section{Status and prospects}

\begin{figure}[!h]
\begin{center}
\begin{overpic}[width=.475\textwidth]{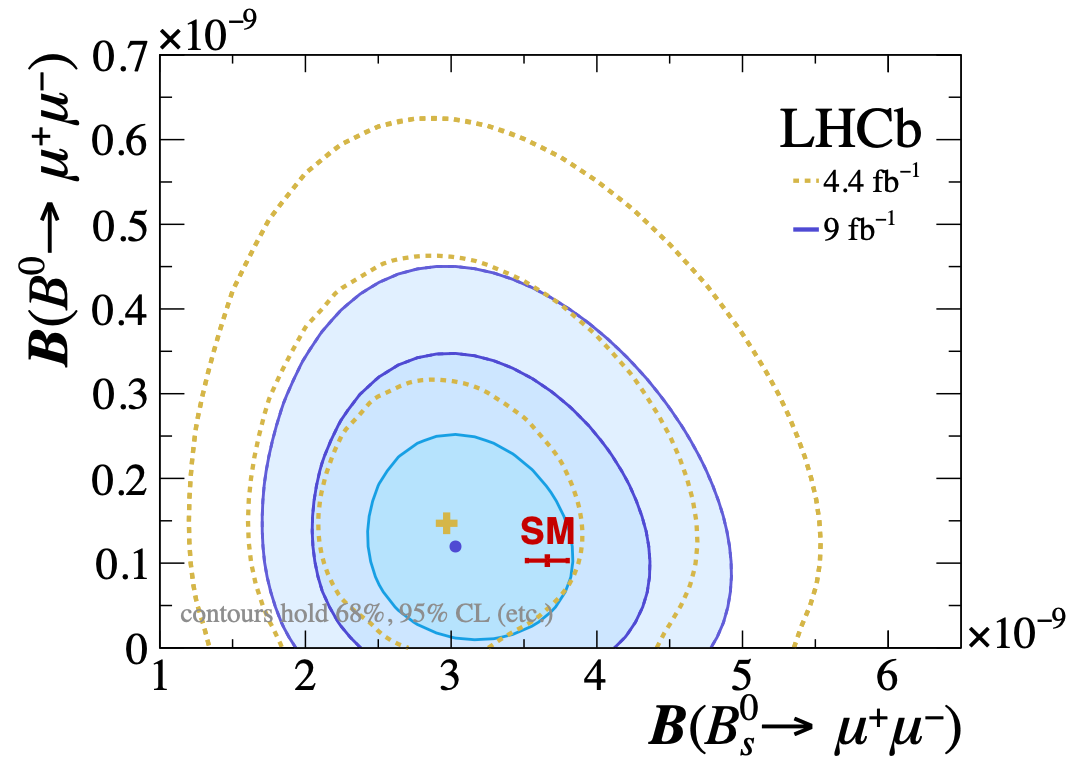}\end{overpic}
\begin{overpic}[width=.455\textwidth]{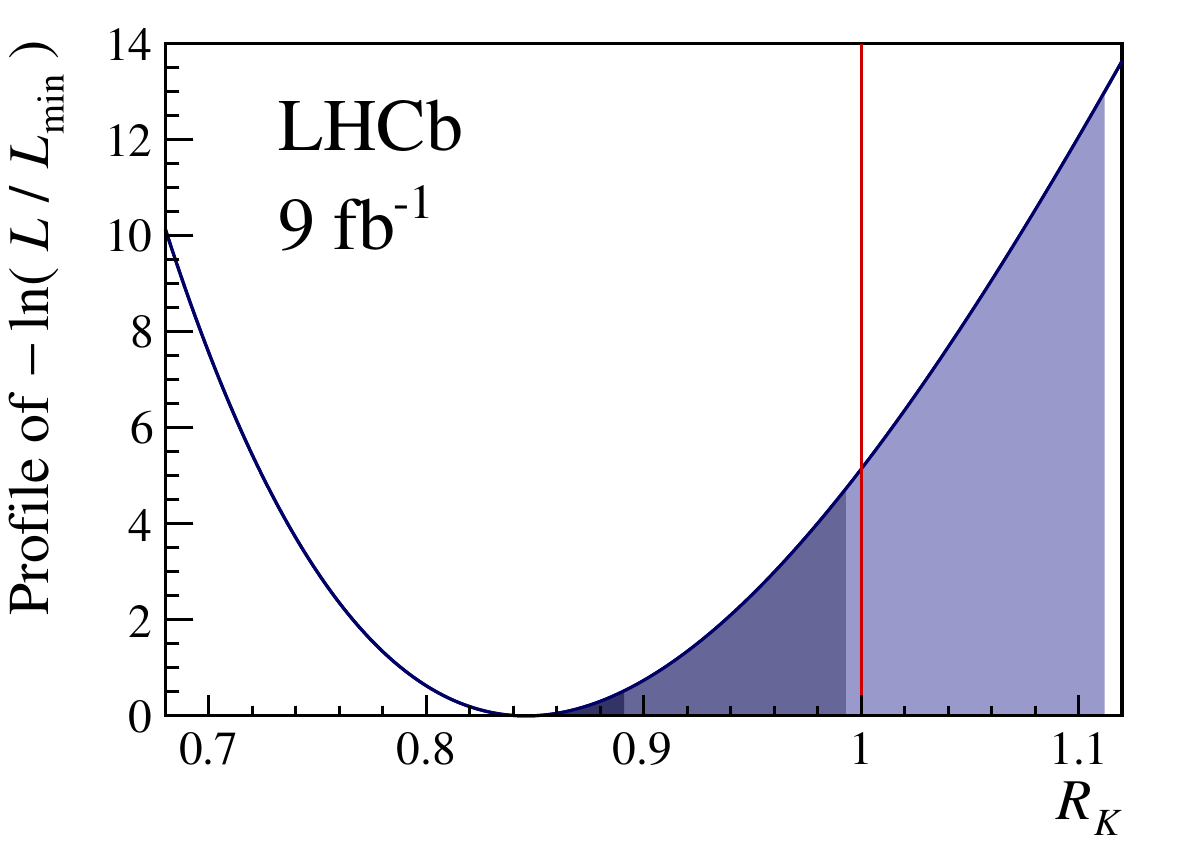}\end{overpic}
\end{center}
\vspace{-.75cm}
\captionof{figure}{Experimental status of \Bsdmumu (left) and \RK (right) after the Moriond 2021 conference. Shown on the left are $1$--$3\,\sigma$ confidence regions (shades of blue) of \Bsmumu and \Bdmumu measurements from LHCb~\cite{LHCb-PAPER-2021-008,LHCb-PAPER-2021-007}, alongside the SM prediction (red). Shown on the right is the likelihood function from the fit to \BKll candidates , profiled in \RK. The extent of the dark, medium and light blue regions show the values allowed for \RK at the $1,\,3,$ and $5\,\sigma$ levels, respectively. The SM prediction is shown as a solid line at $\RK=1$.
}\label{fig:resultsNew}
\end{figure}

The updaded results on \RK and the \Bsdmumu branching fractions are summarised in \Cref{fig:resultsNew}. While the SM prediction for $\mathcal{B}(\Bsdmumu)$ still lies within the $95\% \textrm{ C.L.}$ of the experimental result, the central value is relatively unchanged with respect to the previous LHCb measurement. The same is true for the central value of the \RK result: thanks to the increase in statistics with respect to the previous measurement, it now exhibits $\significance\,\sigma$ tension with respect to the value predicted by the SM~\cite{GinoRKPrediction1Percent,Isidori:2020acz}. Therefore, this \RK measurement represents the first evidence for the violation of LFU in \BKll decays.

Since all presented measurements are dominated by the statistical uncertainty, future updates are crucial to the better understanding of the flavour puzzle. In \Cref{tab:ratiosProspects} are summarised the projected improvements on the precision of the measurements of \RK and $\mathcal{B}(\Bsdmumu)$ in the following years of data taking. The limiting factor in the precision of \RK is the yield of the electron mode, which, assuming the same detector performance as in Run 1, would increase more than $40$ times at the end of Run 4. This would bring the statistical uncertainty on \RK on par with the QED corrections. The error projection on \Bsmumu at future upgrades depends on the assumptions made on the systematic uncertainties, particularly $f_{s}/f_{d}$. A conservative estimate of $4\%$ would imply an uncertainty of $\mathcal{B}(\Bsmumu)$ to be approximately $0.30 \times 10^{-9}$ with $23\,{\rm fb}^{-1}$ and  $0.16 \times 10^{-9}$ with  $300\,{\rm fb}^{-1}$ of data~\cite{LHCb-PII-Physics}.

\begin{table}[!h]
    \captionof{table}{Extrapolation, based on Run 1 results, of statistical uncertainties on \RK, $\mathcal{B}(\Bsmumu)$ (expressed in units of their SM predictions), and the expected electron-mode yield at the end of future detector upgrades. The $b\overline b$ production cross-section is assumed to scale linearly with centre-of-mass energy, and the detector performance is assumed to be unchanged with respect to Run 1. Adapted from Ref.~\cite{LHCb-PII-Physics}.}\label{tab:ratiosProspects}
    \begin{center}
        \begin{tabular}{l | r r r }
            \hline  
            Stat. uncertainty        & $9\,{\rm fb}^{-1}$ & $23\,{\rm fb}^{-1}$ & $300\,{\rm fb}^{-1}$ \\
            \hline  
            \RK     & $4.3\%$    & $2.5\%$                  & $0.7\%$  \\
            \Bsmumu & $12.5\%$   & $8.2\%$                  & $4.4\%$  \\
            \hline
            Yield        & $9\,{\rm fb}^{-1}$ & $23\,{\rm fb}^{-1}$ & $300\,{\rm fb}^{-1}$ \\
            \hline  
            \BKee   & $1120$   & $3300$      & $46000$  \\
        \end{tabular}
    \end{center}
\end{table}

\begin{figure}[!h]
\begin{center}
\begin{overpic}[width=.95\textwidth]{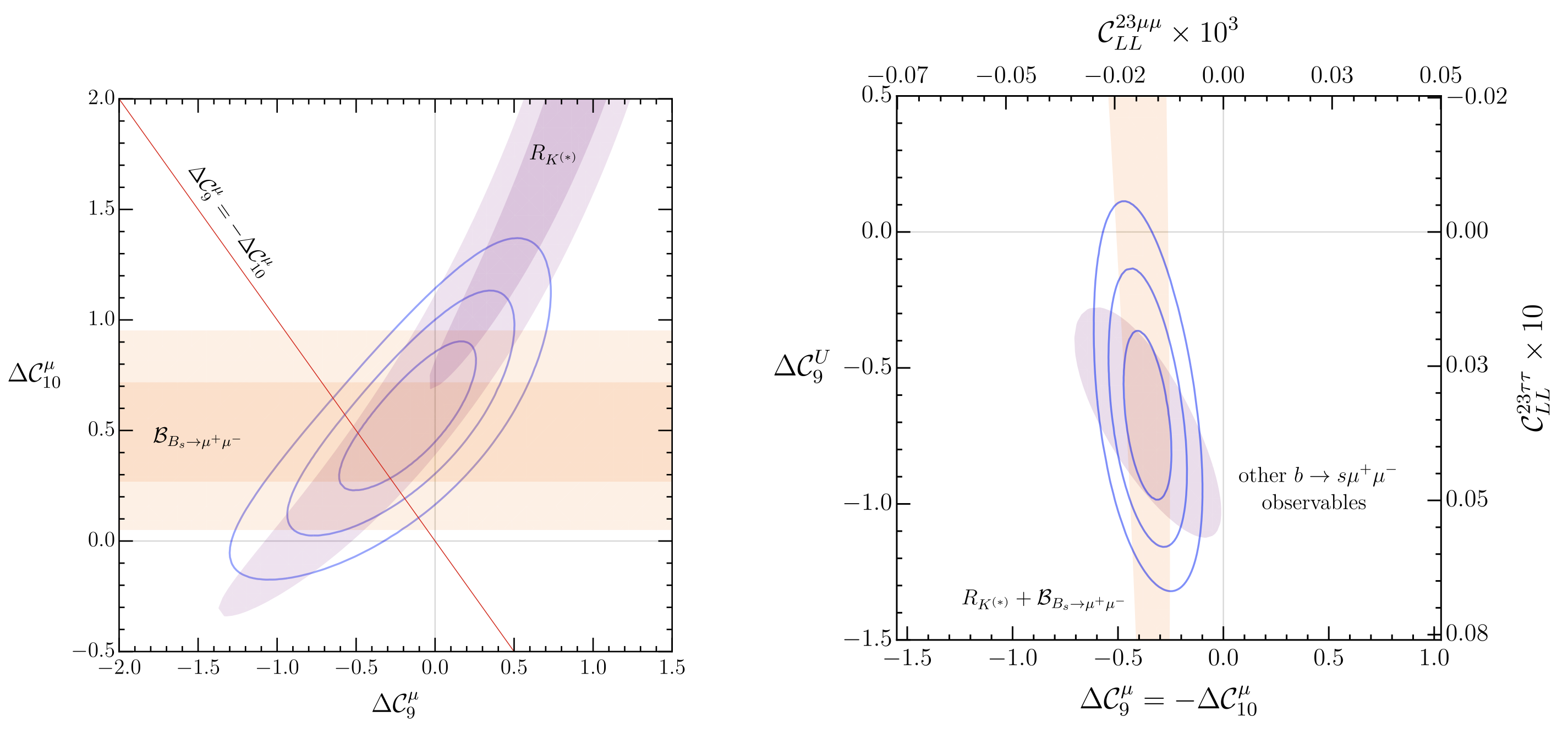}
\put(27.1,12.2){$\star$ {\scriptsize SM}}
\put(77.4,31.8){$\star$ {\scriptsize SM}}
\end{overpic}
\end{center}
\vspace{-.75cm}
\captionof{figure}{EFT constraints from the $\btosll$ anomalies ~\cite{Cornella:2021sby}. Left: Results of the two-dimensional fit to $\Delta C^{\mu}_{9}$ and $\Delta C^{\mu}_{10}$ using clean observables only ($1\sigma$, $2\sigma$ and $3\sigma$ intervals). Also shown are the $1\sigma$ and $2\sigma$ intervals from \RK, \RKst, and \Bsmumu, the latter under the hypothesis $\Delta C^{U}_{9} = 0$. Right: Results of the two-dimensional fit $\Delta C^{\mu}_{10} = -\Delta C^{\mu}_{9}$ vs. $\Delta C^{U}_{9}$ using all \btosll observables. The vertical band shows the result using clean observables only ($1\sigma$ interval), while the ellipse denotes the contribution of all the other observables, estimated using Flavio\cite{Straub:2018kue} ($1\sigma$ interval).}
\label{fig:globalFitsNew}
\end{figure}

The EFT interpretation in terms of shifts in Wilson Coefficients $\Delta C^{\mu}_{i}$ from their SM value, in light of the new results of \RK and \Bsdmumu, is updated in \Cref{fig:globalFitsNew}~\cite{Cornella:2021sby}.
In the plot on the left, the light orange horizontal band highlights the interval of $\Delta C^{\mu}_{10}$ preferred by the combined ATLAS, CMS, and updated LHCb measurements of $\mathcal{B}(\Bsmumu)$, 
while the elongated purple shape marks the regions of $\Delta C^{\mu}_{9}$ and $\Delta C^{\mu}_{10}$ compatible with the measured values of \RK and \RKst \cite{Aaij:2017vbb}.
Thanks to their different sensitivity to the $\Delta C^{\mu}_{i}$ the combination of these two measurement forms the unfilled contours, which are significantly displaced from the SM point. In the plot on the right, the contribution from other \btosll observables, such as the angular ones in the \BKstmumu system ~\cite{Aaij:2020nrf} is included in the $\Delta C_{i}$ fit. While the clean observables presented in this article are insensitive to a flavour-universal redefinition of  $C^{\ell, {\rm SM}}_{9}$ in the SM, ($\Delta C^{U}_{9}$), the observables characterised by less precise theoretical predictions are nonetheless able to provide substantial sensitivity to a shift in $\Delta C^{\mu}_{9}$ and $\Delta C^{\mu}_{10}$, when combined with \RK and $\mathcal{B}(\Bsmumu)$.

In light of the recent results from LHCb, the flavour puzzle has become even more intriguing. 
LHCb will continue to improve the precision on its anomalous flavour measurements, whilst also investigating related observables that could provide complementary information.
Verification from other experiments, such as Belle II, is expected in the near future, so exciting times lay ahead of the particle physics community!

\setboolean{inbibliography}{true}
\bibliographystyle{LHCb}%
\bibliography{main_bibliography,anomalies-theory,LHCb-PAPER}
\end{document}